%% file: k2shrinkage.TeX
\documentclass[Royal,times,sagev]{sagej}
\usepackage{moreverb,url}
\usepackage{booktabs}
\usepackage{verbatim}

\providecommand{\tauprior}{p_{\star}}
\providecommand{\diff}{\mathrm{d}}
\providecommand{\prob}{\mathrm{P}}

\providecommand{\new}[1]{#1}

\begin{document}
\runninghead{R\"{o}ver \& Friede}
\title{Dynamically borrowing strength from another study through shrinkage estimation}
\author{Christian R\"{o}ver\affilnum{1} and Tim Friede\affilnum{1}}
\affiliation{\affilnum{1}University Medical Center G\"ottingen, Department of Medical Statistics, G\"{o}ttingen, Germany.}
\corrauth{Christian R\"{o}ver, University Medical Center G\"ottingen, Department of Medical Statistics, Humboldtallee~32, 37073~G\"ottingen, Germany}
\email{christian.roever@med.uni-goettingen.de}

\begin{abstract}
  Meta-analytic methods may be used to combine evidence from different
  sources of information.  Quite commonly, the normal-normal
  hierarchical model (NNHM) including a random-effect to account for
  between-study heterogeneity is utilized for such analyses.  The same
  modeling framework may also be used to not only derive a combined
  estimate, but also to borrow strength for a particular study from
  another by deriving a \emph{shrinkage estimate}.  For instance, a
  small-scale randomized controlled trial could be supported by a
  non-randomized study, e.g. a clinical registry.  This would be
  particularly attractive in the context of rare diseases.  We
  demonstrate that a meta-analysis still makes sense in this extreme
  case, effectively based on a synthesis of only two studies, as
  illustrated using a recent trial and a clinical registry in
  Creutzfeld-Jakob disease.  Derivation of a shrinkage estimate within
  a Bayesian random-effects meta-analysis may substantially improve a
  given estimate even based on only a single additional estimate while
  accounting for potential effect heterogeneity between the studies.
  \new{Alternatively, inference may equivalently be motivated via a model specification that does not require a common overall mean parameter but considers the treatment effect in one study, and the difference in effects between the studies.}
  The proposed approach is quite generally applicable to combine
  different types of evidence originating e.g. from meta-analyses or
  individual studies. An application of this more general setup is
  provided in immunosuppression following liver transplantation in
  children.
\end{abstract}

\maketitle                   






\section{Introduction}
  In clinical research of orphan diseases, one of the major problems
  is often the recruitment of a sufficient number of subjects to
  perform a meaningful clinical trial.  Examples include neuromyelitis
  optica \cite{ChatawayFriede2016}, myocarditis
  \cite{MessroghliEtAl2017}, and Creutzfeld-Jakob disease (CJD)
  \cite{VargesEtAl2017}.  In such cases, it may be possible to gain
  some power by using more sophisticated trial designs, and it is
  often desirable to be able to formally utilize additional
  information external to the actual trial, which may be implemented
  via the use of informative prior distributions in the eventual
  analysis \cite{AbrahamyanEtAl2016}. The external information could
  be in the form of related studies or elicited expert opinion
  \cite{HampsonEtAl2014}. For instance, in the context of a
  small-scale randomized controlled trial in idiopathic nephrotic
  syndrome in children, a rare condition, Thall \textit{et~al.}
  \cite{ThallEtAl2017} recently proposed the elicitation of expert
  opinions on response probabilities based on the bins-and-chips
  approach \cite{JohnsonEtAl2010}.

  When considering external evidence, the obvious danger is that a too
  simplistic approach may lead to a ``na\"{i}ve'' pooling of initially
  separate data.  For example, while data from non-randomized studies
  (e.g. clinical registries) may undoubtedly contribute complementing
  information to a randomized clinical trial, one may want to prevent
  a complete mixing of both types of data, which would in a sense also
  invalidate the original randomization. Rather it seems desirable to
  stratify the analysis for the different sources by explicitly
  allowing for potential heterogeneity between data sets, which then
  implicitly downweights the impact on one another.  Here the weights
  depend on the observed similarity of estimates, also known as
  \emph{dynamic borrowing} of information \cite{VieleEtAl2014}.  The
  eventual analysis then may refer explicitly to the outcome of the
  randomized trial, and not to some overall average, as generally more
  weight is placed on evidence from randomized controlled trials.

  A simple approach originally proposed by Pocock \cite{Pocock1976} was
  recently implemented by Schoenfeld
  \textit{et~al.} \cite{SchoenfeldZhengFinkelstein2009}, who
  investigated the use of adult data to support the analysis of a
  paediatric trial, and who utilized a variance component of known
  (elicited) magnitude to account for heterogeneity between the two
  studies' estimates.  A closely related approach is implemented in
  the normal-normal hierarchical model (NNHM) that is commonly
  utilized in random-effects meta-analysis; the difference essentially
  is that heterogeneity is treated as an unknown for which a prior
  distribution may be specified.  Technically, inference on the study
  of primary interest is done by investigating the corresponding
  \emph{shrinkage estimate}. The contribution of information from
  additional studies then may readily be evaluated by considering the
  corresponding \emph{meta-analytic-predictive (MAP)} prior
  \cite{SchmidliEtAl2014,WandelNeuenschwanderRoeverFriede2017}.  The
  NNHM is readily generalized (and in fact most commonly used) for
  combining more than two studies; such an approach may e.g. be used
  to extrapolate information from early-phase studies in the approval
  process \cite{WandelNeuenschwanderRoeverFriede2017}.  In the case of
  two studies, the NNHM can also be shown to be to some extent
  equivalent to a similar, more general model specification as we will
  explain below.

  While the interpretation of parameters within the familiar NNHM
  context is straightforward and with the inclusion of an unknown
  heterogeneity parameter it is intended that evidence from separate
  studies is sufficiently loosely connected to provide a robust
  estimation framework, it is not obvious to what extent this approach
  actually improves estimates in the extreme case of only two studies
  or more generally two data sources. Here we develop a suitable
  statistical hierarchical model to include two sources of data, e.g.,
  two studies or meta-analyses.  Within the proposed model we describe
  a shrinkage estimator and inference methods including posterior
  predictive $p$-values. Furthermore, the value of this approach in
  the particular case of only two studies is evaluated in simulations.
	
  The manuscript is organized as follows. In the next section we
  present the statistical model and, in particular, shrinkage
  estimation and inference. The following sections are dedicated to a
  simulation study investigating the operating characteristics of the
  proposed methodology and an application in CJD\@. Motivated by a
  meta-analysis investigating the effect of immunosuppression in
  paediatric liver transplantation patients, we extend the
  shrinkage applications to more general settings considering two data
  sources, e.g. two meta-analyses, rather than two studies. Finally,
  we close with some conclusions and a brief discussion.

\section{Statistical model and shrinkage estimation}
\subsection{The normal-normal hierarchical model}\label{sec:nnhm}
  The most commonly used model for random-effects meta-analysis is the normal-normal
  hierarchical model (NNHM). This model is applicable for the joint
  analysis of several ($k$) real-valued effect measurements~$y_i$
  that have individual standard errors~$\sigma_i$ associated ($i=1,\ldots,k$). Here it is
  assumed that each observation~$y_i$ is a noisy measurement of an
  underlying true value~$\theta_i$ with a normally distributed offset
  whose magnitude is given through the (known) standard error~$\sigma_i$:
  \begin{equation}\label{eqn:nnhm-1}
    y_i|\theta_i \; \sim \; \mathrm{N}(\theta_i, \sigma_i^2) \mbox{.}
  \end{equation}
  The $\theta_i$ then may be more or less similar across measurements;
  at the study level, a certain amount of \emph{heterogeneity} is
  anticipated by introducing another variance component~$\tau$ and assuming
  \begin{equation}\label{eqn:nnhm-2}
    \theta_i|\mu,\sigma \; \sim \; \mathrm{N}(\mu, \tau^2)
  \end{equation}
  where the overall effect~$\mu$ and heterogeneity~$\tau\geq 0$ are
  unknown. If $\tau=0$, the model simplifies to the
  \emph{fixed-effect} model, in which $\theta_1=\cdots=\theta_k=\mu$,
  but in general this is a \emph{random-effects} model
  \cite{HedgesOlkin,HartungKnappSinha,BorensteinEtAl}.  In the
  following, we will mostly be concerned with the special case of
  analyzing only $k=2$ studies.  Note that while classically, in the
  meta-analysis context, the $y_i$ usually originate from different
  \emph{studies}, more generally these may also be \emph{estimates} of
  other kinds, e.g., estimates from meta-analyses.

\subsection{Shrinkage estimation}
  Quite commonly, the main interest lies in determining the
  \emph{overall effect}~$\mu$. When the aim of the analysis is to
  provide a basis for planning a new study, it may also be of interest
  to \emph{predict} a future outcome~$\theta_{k+1}$.  
  In some cases, however, it is of interest to derive an updated
  estimate for a particular ($i$th) study effect~$\theta_i$, which
  is informed by the remaining studies under consideration. 

  If the heterogeneity~$\tau$ was zero, then the model would reduce to
  the fixed-effect model, and all estimates~$y_i$ would effectively
  relate to the same parameter ($\theta_i=\mu$), which may then be
  jointly estimated by simply averaging the estimates~$y_i$ (with
  ``inverse variance'' weights).  If the heterogeneity appears to be
  close to zero, then the model will behave similarly to the
  fixed-effect model, and all estimated study effects~$\theta_i$ will
  be ``shrunk'' towards the estimated overall mean~$\mu$ to some
  degree.  If on the other hand the heterogeneity is large, then there
  is very little to be learned from one estimate ($y_i$) about another
  parameter ($\theta_j$, $j \neq i$), and different estimates only
  provide very little support to one another. Effectively, this
  results in more or less \emph{shrinkage} towards the overall
  mean~$\mu$, depending on the apparent heterogeneity in the data
  \cite{BDA3rd,WandelNeuenschwanderRoeverFriede2017}.  This
  \emph{shrinkage estimation} of study-specific means~$\theta_i$,
  which is also known as \emph{best linear unbiased prediction (BLUP)}
  in a frequentist framework
  \cite{RaudenbushBryk1985,Robinson1991,Viechtbauer2010}, will be our
  focus in the following.  When the heterogeneity~$\tau$ is assumed
  fixed and known (and an improper uniform prior for~$\mu$ is used),
  then the frequentist and Bayesian approaches lead to identical mean
  effect ($\mu$) \cite{FriedeRoeverWandelNeuenschwander2017a} as well
  as shrinkage ($\theta_i$) estimates
  \cite{RaudenbushBryk1985,Roever2017}; in general, however, these are
  different.

\subsection{The Bayesian approach to meta-analysis}
  The inference problems within the NNHM may be approached using
  frequentist or Bayesian methods
  \cite{HedgesOlkin,HartungKnappSinha,BorensteinEtAl,SpiegelhalterEtAl,Roever2017}. A
  Bayesian approach has proven especially useful in cases where
  large-sample asymptotics do not apply \cite{RoeverFriede2018}, e.g.,
  for the analysis of few studies
  \cite{FriedeRoeverWandelNeuenschwander2017a} or even only two
  studies \cite{FriedeRoeverWandelNeuenschwander2017b}.  Here, we will
  follow a Bayesian approach and investigate its properties in some
  more detail.

  Within the NNHM we have several unknowns; firstly the study-specific
  effects~$\theta_i$, whose hyperprior is given
  through~(\ref{eqn:nnhm-2}). For the overall mean effect~$\mu$ it is
  often convenient to use a non-informative (improper) uniform
  prior. The heterogeneity~$\tau\geq 0$ determines the expected
  variability between individual studies; depending on the measurement
  scale of the considered effects, in practical applications a
  plausible upper bound can usually be specified.  Half-normal (HN)
  priors (e.g. with scale parameters $0.5$ or $1.0$) have proven
  useful for example in the context of logarithmic odds-ratio (log-OR)
  endpoints
  \cite{SpiegelhalterEtAl,Gelman2006,FriedeRoeverWandelNeuenschwander2017a}.
  An analogous reasoning similarly applies for many log-transformed
  endpoints, like relative risks or hazard ratios; if different
  studies are considered unlikely to differ by more than a certain
  \emph{factor}, then one can usually translate this into a prior
  specification for the heterogeneity~$\tau$ on the logarithmic scale
  \cite{SpiegelhalterEtAl,Roever2017}.

  The heterogeneity~$\tau$ is usually considered a nuisance parameter,
  while the primary interest is in inferring the overall effect
  ($\mu$), a prediction ($\theta_{k+1}$), or a shrinkage estimate
  ($\theta_i$).  Within the Bayesian framework, shrinkage estimation
  may be motivated in two different ways; the
  \emph{meta-analytic-combined (MAC)} approach simply considers the
  shrinkage estimate as one of the parameters in the NNHM model, where
  all~$k$ studies are analyzed jointly. The
  \emph{meta-analytic-predictive (MAP)} approach on the other hand
  considers the same problem sequentially: first, all but the $i$th
  study are analyzed, and the derived posterior (predictive)
  distribution then constitutes the prior for the analysis of the
  $i$th study. Both approaches can be shown to be equivalent and lead
  to identical results \cite{SchmidliEtAl2014}, but the MAP approach
  allows to explicate the information `borrowed' from the additional
  estimates via the MAP~prior.  Technically, inference requires
  integration over the parameters' posterior distribution
  \cite{BDA3rd,SpiegelhalterEtAl}, e.g. to derive the relevant
  marginal posterior distribution for shrinkage estimation
  ($p(\theta_i|y_1,y_2,\sigma_1,\sigma_2)$).  Computations for
  inference within the NNHM may be performed in \textsf{R} using the
  \texttt{bayesmeta} package \cite{bayesmeta,Roever2017}.  In the
  following, the shown estimates will be posterior medians, and
  credible intervals are determined as shortest posterior intervals.

\subsection{Posterior predictive $p$-values}
  Posterior predictive $p$-values are conceptually closely related to
  ``classical'' $p$-values, and were originally developed in the
  context of model checking
  \cite{Meng1994,GelmanMengStern1996,BDA3rd}. The definition is
  relatively straightforward; like a classical $p$-value, it is based
  on a null hypothesis~$H_0$ and a pre-specified (``test\mbox{-}'')
  statistic or ``discrepancy variable''~$T(\cdot)$, which is a
  function of the data. The statistic~$T$ is (as usual) defined so
  that it is sensitive to deviations from the null hypothesis. The
  realised statistic value~$T(y)$ then is determined for the present
  data set~$y$. In order to judge whether the statistic value is
  ``sufficiently extreme'' to constitute evidence against the null
  hypothesis, it is compared against its \emph{posterior predictive
    distribution, conditional on $H_0$}.

  Similarly to the usual $p$-values, this means a comparison against
  values of the statistic amongst data sets that might have occurred
  \emph{conditioning on the observed data as well as the null
    hypothesis}. Technically, posterior predictive $p$-values are
  often easily computed using Monte Carlo sampling, which here means
  first drawing parameter values from the intersection of the
  parameters' posterior distribution and null hypothesis, then drawing
  a data set~$y^\star$ from the corresponding predictive distribution,
  and determining the statistic value~$T(y^\star)$. Repeated sampling
  then allows to explore the relevant distribution of statistic values
  and eventually compute $p$-values via the corresponding tail
  probabilities. While posterior predictive $p$-values generally do
  not follow a uniform distribution under the null hypothesis, the
  deviation is usually on the conservative side
  \cite{Meng1994,Gelman2013}.

  The test statistic to be used needs to be pre-specified. For
  instance, an obvious choice for the overall effect~$\mu$ may be the
  posterior probability of a non-beneficial effect, i.e.,
  \begin{equation}
     T(y) \;=\; \mathrm{P}(\mu > 0 \,|\, y)\mbox{.}
  \end{equation}
  The null hypothesis then is usually specified for a certain
  parameter as one- or two-sided. Accordingly, the test statistic's
  relevant distribution (or the sampling scheme, in case of MCMC
  computation), as well as the statistic's rejection region are
  affected. Computation of posterior predictive $p$-values is also
  implemented in the \texttt{bayesmeta} \textsf{R} package
  \cite{Roever2017}.

\subsection{The reference model as an alternative variation of the NNHM}\label{sec:altnnhm}
  When meta-analyzing a pair of estimates, the common NNHM may
  sometimes be hard to motivate, as an exchangeable model of both
  estimates~$\theta_i$ centered around a common mean value~$\mu$ may
  seem inappropriate.  Consider for example the joint analysis of
  randomized and observational data; reference to a common mean
  parameter~$\mu$ or identical variances~$\tau^2$ may be
  counterintuitive in such a case.  An ``asymmetric'' treatment of
  both estimates in terms of a ``reference'' estimate and a
  ``secondary'', related observable with an uncertain amount of offset
  associated may seem more appealing.  It is possible to formulate a
  slight variation of the NNHM following this second approach, which
  may seem more realistic, and for which one can then show that both
  approaches are equivalent \emph{as far as shrinkage estimates are
    concerned}.  In the following, we will refer to this model
  variation as the \emph{reference model}.

  Suppose that the prior for the effect~$\mu$ in the NNHM is given by
  an (improper) uniform distribution, and that the heterogeneity prior
  is defined through a density~$\tauprior(\tau)$. Then the model
  variation is defined as follows; for the observables $y_i$ we assume
  \begin{eqnarray}\label{eqn:altnnhm-1}
    y_i|\vartheta_i & \sim & \mathrm{N}(\vartheta_i, \sigma_i^2) \mbox{,}
  \end{eqnarray}
  which so far is analogous to the NNHM setup. At the next hierarchy
  level, we then specify
  \begin{eqnarray}\label{eqn:altnnhm-2}
    \vartheta_1|\alpha,\beta & \sim & \mathrm{N}(\alpha, 0) 
      \qquad \mbox{(i.e., }\vartheta_1=\alpha \mbox{)} \mbox{,}\\
    \vartheta_2|\alpha,\beta & \sim & \mathrm{N}(\alpha, \beta^2) \mbox{.}
  \end{eqnarray}
  where the ``effect'' parameter~$\alpha$ again has an improper
  uniform prior and the variance component~$\beta$ now has a prior
  density given by
  $\frac{1}{\sqrt{2}}\tauprior\bigl(\frac{\beta}{\sqrt{2}}\bigr)$.
  The parameter $\beta$ hence has a prior that is scaled by a factor
  of~$\sqrt{2}$ relative to $\tau$, which corresponds to a factor~$2$
  difference for the \emph{squared} parameters (the variances).

  The reference model parametrisation of the problem is different here
  in that the two observables $y_i$ are treated asymmetrically. The
  first one ($y_1$) measures the parameter~$\alpha$ (the reference)
  ``directly'', while the second one ($y_2$) includes an additional
  offset with variance~$\beta^2$. The variance component~$\beta$ again
  implements the heterogeneity between first and second observable,
  but in a slightly different manner than in the original NNHM\@.
  While the parameterizations are different, the associated shrinkage
  estimates (for $\theta_i$ or $\vartheta_i$) are identical, as is
  shown in detail in the Appendix.  Since $\vartheta_1=\alpha$, the
  shrinkage estimate for $\vartheta_1$ is identical to an estimate of
  $\alpha$ in this context. The NNHM's heterogeneity ($\tau$) prior
  needs to be re-scaled by a factor of $\sqrt{2}$ to yield the
  corresponding $\beta$~prior.  Note, however, that the equivalence
  only holds for the case of $k=2$ estimates, and an (improper)
  uniform effect prior; for other cases, the model would need to be
  adapted accordingly.

  As has been pointed out by Neuenschwander \textit{et~al.}
  \cite{NeuenschwanderEtAl2016b}, the model may also be regarded as a special
  case of Pocock's bias model, or the model underlying the commensurate
  prior. 
  In both instances, for the case of $k=2$ studies, the discrepancy
  between the two underlying parameters (here: $\vartheta_1$
  and~$\vartheta_2$) is also modeled via a variance parameter
  analogous to~$\beta^2$ above.
  The connection is made somewhat differently in the \emph{power
    prior} model \cite{IbrahimChen2000}, where the external data are
  downweighted via an exponential parameter between~$0$ and~$1$ that is applied to
  part of the likelihood function. For a given $\tau$ (or~$\beta$)
  value, the approaches are again identical when the exponential parameter is set
  to be $\bigl(2\frac{\tau^2}{\sigma_2^2}+1\bigr)^{-1}$ or
  $\bigl(\frac{\beta^2}{\sigma_2^2}+1\bigr)^{-1}$.

\section{Dependency of the shrinkage estimate on the observed heterogeneity}
  In the following, we investigate the effect of varying the input data
  on the resulting shrinkage estimates.  
  The setup is similar to the
  one also adopted in the subsequent simulation study; we consider the
  case of two estimates ($y_1$ and $y_2$) with standard errors
  $\sigma_1=0.8$ and $\sigma_2=0.2$, and we assume a uniform prior
  for~$\mu$ and a half-normal (HN) prior with scale~$0.5$ for the
  heterogeneity~$\tau$. We set $y_1=0$ and then vary the difference
  between the estimates ($y_2-y_1$), which is in a sense also the
  ``observed heterogeneity'' in the data. Then we derive the shrinkage
  estimate for the first parameter~$\theta_1$.

  \begin{figure*}[ht]
    \begin{center} 
      \includegraphics[width=0.66\linewidth]{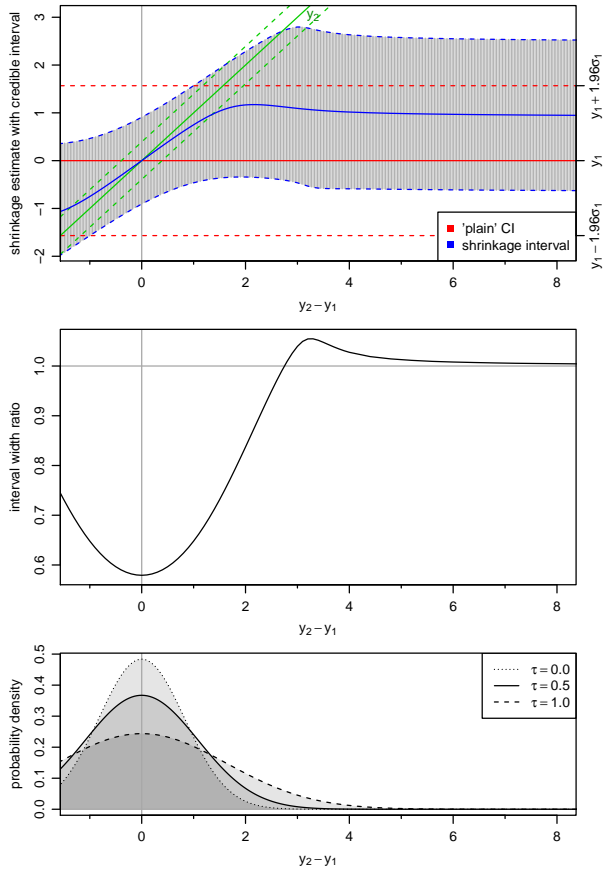} 
      \caption{\label{fig:comparison} Effect of varying the difference
        between quoted estimates ($y_2-y_1$) on the first shrinkage
        estimate (for~$\theta_1$).  In the top row, one can see how
        the interval itself varies relative to the ``plain'' interval
        ($y_1 \pm 1.96 \sigma_1$, red lines); the second estimate~($y_2$) 
        and its corresponding CI are shown in green.
        The second row shows the ratio of
        interval lengths, and the bottom row shows the probability density
        of the actualized difference for selected values
        of~$\tau$. The heterogeneity prior used here for the analysis
        was half-normal (HN) with scale~$0.5$.}
    \end{center}
  \end{figure*}

  Figure~\ref{fig:comparison} (top panel) illustrates the effect on
  the shrinkage estimate and the corresponding $95\%$~credible
  interval. One can see how the estimate (posterior median
  of~$\theta_1$) moves (mostly) in concordance with the second
  estimate ($y_2$) and that the resulting interval is narrowest when
  $y_1$ and $y_2$ are in close agreement. For larger differences, the
  estimated heterogeneity increases, less borrowing of information
  takes place, the interval widens and the estimate of~$\theta_1$ is
  less attracted towards $y_2$. Eventually the shrinkage interval
  exhibits a certain degree of robustness and barely changes with
  increasing difference.  This robustness feature may be explained by
  the fact that implicitly the meta-analysis is equivalent to an
  analysis of the first study using the MAP-prior based on the second
  study \cite{SchmidliEtAl2014}. The prior derived via the
  hierarchical model from the first study then is rather vague and
  heavy-tailed, leading to the robust behaviour
  \cite{OHaganPericchi2012}.

  The middle panel shows that the shrinkage interval is shorter than
  the ``plain'' interval ($y_1 \pm 1.96 \sigma_1$) when the estimates
  $y_1$ and $y_2$ are similar, that it may also get wider in some
  cases, but that its width eventually is bounded. The bottom panel
  shows the probability distribution of the difference $y_2-y_1$
  (which has variance $\sigma_1^2+\sigma_2^2+2\tau^2$) for several
  selected values of~$\tau$. Based on the assumptions, the absolute
  difference is unlikely to exceed a value of, say, $|y_2-y_1|=4$, and
  so the probable cases essentially are those in the left half of the
  plot.

  The scenario shown here is where we would in fact expect the
  greatest gain from considering the second estimate ($y_2$) in
  estimating $\theta_1$, since the second estimate's error is much
  smaller than the first ($\sigma_2 \ll \sigma_1$).  The figure looks
  qualitatively similar if we match or reverse the relative magnitudes
  of the standard errors $\sigma_1$ and $\sigma_2$, and also if we use
  a wider heterogeneity prior, but in those cases there is less
  information to be borrowed and hence less ``shrinkage'' taking
  place.

\section{Simulation study}
\subsection{Setup}
  The simulations shown in the following are based on the 
  NNHM, and since binary endpoints are very common in meta-analysis applications \cite{DaveyEtAl2011},
  the
  setup is motivated by a scenario featuring a log-OR endpoint.  If a
  study of size~$n_i$ results in a contingency table as an outcome,
  this may be converted into a log-OR that is associated with an
  approximate standard error of $\sigma_i=\frac{4}{\sqrt{n_i}}$ \cite{Roever2017}. A similar formula applies e.g.\ for logarithmic hazard ratios (log-HRs) from a survival analysis with respect to the event counts \cite{SpiegelhalterEtAl}.
  In
  the following we will consider combinations of ``small'', ``medium''
  and ``large'' studies of sizes $n_i\in\{25, 100, 400\}$,
  corresponding to standard errors of $\sigma_i\in\{0.8, 0.4, 0.2\}$.
  The true mean effect~$\mu$ is (arbitrarily) fixed at zero.
  Analysis of a pair of studies will be based on a uniform
  prior for the effect~$\mu$, and a half-normal (HN) prior for the
  heterogeneity~$\tau$.  Heterogeneity values in the range $0.5$--$1.0$
  may be considered as \emph{fairly high} and above~1.0 as
  \emph{fairly extreme} \cite{SpiegelhalterEtAl}.  A prior scale
  parameter of~$0.5$ already is a conservative choice, but in addition
  we also investigate the use of a HN($1.0$) prior
  \cite{SpiegelhalterEtAl,FriedeRoeverWandelNeuenschwander2017a}.  The
  true heterogeneity values in the simulation will be varied among
  $\tau\in\{0.0,0.1,0.2,0.5,1.0,2.0\}$ in order to check performance
  conditional on particular~$\tau$ values. Similarly, we investigate
  the \emph{marginal} performance by drawing $\tau$ according to the
  specified prior distribution.  The primary interest will be in the
  first of the two studies ($i\!=\!1$) and especially the shrinkage
  estimate of its study-specific effect~$\theta_1$.  The number of
  simulations for each scenario is $10\,000$.

  We can compare the resulting precision by comparing the
  $95\%$~shrinkage interval width~$\delta_i$ with the original
  confidence interval width and considering the \emph{relative width}
  $q_i=\frac{\delta_i}{2\times 1.96 \sigma_i}$.  Assuming that
  standard errors scale with~$n_i^{-0.5}$, we can then estimate the
  approximate \emph{gain in effective sample size} as $q_i^{-2}-1$.  For
  example, if the shrinkage interval is only half as wide as the
  original interval, this precision gain corresponds to a roughly
  four-fold ($300\%$) increase in sample size.  If the interval is
  $q_i\!=\!90\%$ as wide, then this corresponds to an approximate
  $q_i^{-2}-1 = 23\%$ increase.
  \textsf{R}~code to reproduce the simulations is included in the supplement.


\subsection{Coverage}
  Table~\ref{tab:shrinkCover} illustrates the coverage of shrinkage
  intervals for the effect~$\theta_1$ for different combinations of
  study sizes ($n_1$, $n_2$), heterogeneity values ($\tau$) and
  heterogeneity priors (scales $0.5$ and $1.0$).  The columns marked by an
  asterisk~($\ast$) correspond to the ``marginal'' simulations in
  which heterogeneity~$\tau$ is not fixed, but varied according to the
  specified prior distribution.  Coverages are close to or above the
  nominal $95\%$~level, except if heterogeneity approaches a~priori
  improbable large values. For the simulations in which $\tau$ is
  drawn from its prior distribution, we know that \emph{by
    definition} the coverage would be exactly~$95\%$ \emph{if} the
  effect $\mu$ was also drawn from its prior \cite{GneitingEtAl2007}. Since the effect prior
  is improper and $\mu$ was arbitrarily fixed at zero for the
  simulations, this only holds approximately here.

  \addtolength{\tabcolsep}{-1.0mm}
  \input{shrinkCover.tex}

\subsection{Interval length and effective sample size gain}
  Table~\ref{tab:shrinkWidth} shows the mean lengths of shrinkage
  intervals relative to the original (``plain'') confidence interval
  based on $y_1$ and $\sigma_1$ alone (which has width $2 \! \times \!
  1.96\sigma_1$).  While we have seen in the previous section that
  intervals may be shorter or longer in certain cases, here we see
  that on average the shrinkage intervals are always shorter than the
  plain intervals. As expected, the gain is largest if the study under
  consideration is small relative to the additional evidence
  ($n_1<n_2$, $\sigma_1>\sigma_2$), and if heterogeneity is
  low. Assuming a wider heterogeneity prior also reduces the amount of
  borrowing of information and leads to wider intervals.

  \input{shrinkWidth.tex}

  The gain in precision may approximately be translated to an
  equivalent gain in effective sample size (as expressed through
  the~$q_i$ introduced above).  The average gain is shown in
  Table~\ref{tab:shrinkGain}. This relative gain in information may be
  substantial and is most pronounced if $n_1$~is small relative
  to~$n_2$. For example, for the HN($0.5$) heterogeneity prior and
  $n_1=25$, we can expect a gain of at least one third across all
  scenarios, and even a gain of more than $100\%$ is well achievable
  in certain cases. When averaging over the heterogeneity prior, i.e.,
  if we assume the prior to accurately reflect the probability
  distribution for~$\tau$, we can expect a gain of more than $50\%$
  for the cases where $n_1=25$ and more than $100\%$ when in addition
  $n_2>n_1$.

  \input{shrinkGain.tex}

\subsection{Fraction of shortened intervals}
  While there is a gain \emph{on average}, the shrinkage intervals
  may in some cases also turn out wider than the original interval.
  Table~\ref{tab:shortFraction} shows the percentages of intervals
  showing a smaller width. In a majority of cases, we can expect
  a shorter interval, the exceptions are again cases where the
  heterogeneity is large, or the second study is small.
 
  \input{shortFraction.tex}

\subsection{Implications for practical application}
  The previous sections illustrate the process of shrinkage estimation
  within the NNHM framework and investigate the potential
  benefits. Across a range of realistic settings, the method exhibits
  sensible and robust behaviour, and despite the seemingly
  pathological outset of synthesizing only two estimates, the expected
  information gain may still be substantial. In the following, we will
  illustrate the approach by applying it in two examplary cases, onc
  based on two \emph{studies} (one randomized, one observational), and
  one based on two \emph{estimates} from meta-analyses of different
  types of studies.

\section{An application in Creutzfeld-Jakob disease}
  With a prevalence of $1$ in $1\,000\,000$ \cite{CjdFactSheet} and an
  incidence of $1.5$ per million and year \cite{WakapEtAl2016},
  Creutzfeld-Jakob disease (CJD) is clearly a rare disease by any
  standard. In a recent systematic review, Unkel
  \textit{et~al.} \cite{UnkelEtAl2016} identified a number shortcomings
  in the methodologies applied in clinical studies conducted in CJD
  and advocated the use of innovative statistical methodology
  including evidence synthesis approaches.

  Varges \textit{et~al.} \cite{VargesEtAl2017} studied the use of
  doxycycline, an antiprion agent, in early CJD. They conducted a
  double-blind randomised placebo-controlled trial that failed to
  recruit the originally planned number of patients and was terminated
  prematurely with only $n=12$ patients ($7$ on doxycycline and $5$ on
  placebo). Additionally, data were available from an observational
  study of $n=88$ patients including $55$ patients who received
  doxycycline. The primary endpoint was all-cause mortality which was
  analyzed using Cox proportional hazard regressions. In the case of
  the randomized controlled trial the model included only the factor
  treatment as independent variable whereas the analysis of the
  observational data in addition was stratified by propensity
  scores. The observed log hazard ratios (standard errors) were
  $-0.173$ ($0.631$) and $-0.499$ ($0.249$) in the randomized
  controlled trial and the observational study, respectively. Varges
  \textit{et~al.} performed a random-effects meta-analysis to estimate
  the overall (pooled) effect $\mu$ using standard frequentist
  methodology. They reported a combined hazard ratio of $0.633$ with
  95\% confidence interval of $(0.402; 0.999)$.

\begin{figure}[ht]
    \begin{center}
      \includegraphics[width=0.8\linewidth]{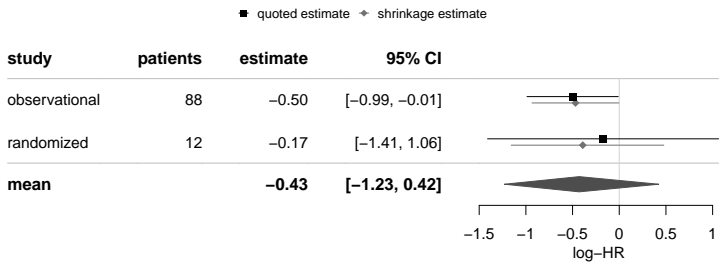} 
      \caption{\label{fig:cjd-example} Forest plot for the CJD~example
        (log-HR outcome). The shrinkage interval for the log-HR based on
        randomized evidence here is [$-1.16, 0.48$], spanning only two
        thirds of the original confidence interval width.}
    \end{center}
  \end{figure}

  Now suppose primary interest was in the `randomized' effect, but one
  is willing to utilize external observational evidence as supporting
  information. We may now apply the shrinkage estimation
  approach. Figure~\ref{fig:cjd-example} shows the estimated
  logarithmic hazard ratios based on obervational and randomized data
  along with the derived mean estimate~($\mu$).  The two shrinkage
  estimates are also shown next to the original (quoted)
  estimates. For the randomized trial, the updated credible interval
  covers the range of \mbox{[$-1.16, 0.48$]} and is only $66\%$ as
  wide as the original interval. This amount of shrinkage implies a
  gain in effective sample size of~$129\%$, i.e., this corresponds to
  more than a doubling of the original sample size from $12$~patients
  to an `effective number' of some $27$~patients.  For the randomized
  patients' shrinkage estimate, we then obtain a posterior probability
  of a non-beneficial effect of $\prob(\theta_\mathrm{rand.} \!>\!
  0\,|\,y)=0.16$. The associated (one-sided) posterior predictive
  $p$-value is similar, with~$p=0.13$.

  From two studies there is only very little to be learned about the
  between-study heterogeneity~$\tau$
  \cite{FriedeRoeverWandelNeuenschwander2017b}. The prior median
  heterogeneity was at~$0.34$, which a~posteriori is slightly reduced
  to~$0.28$; the posterior $95\%$~quantile is at~$0.85$ instead
  of~$0.98$.  Note that while the estimates for the overall mean and
  the shrinkage estimate do not differ much in this particular case,
  their interpretations are quite different. The \textsf{R}~code to
  reproduce the calculations for this example is provided in the
  appendix.

\section{Beyond two studies: more general shrinkage applications} \label{sec:generalization}
  So far we have described shrinkage estimation mostly in terms of
  ``studies'' and corresponding parameter estimates. However, the
  method may be applied more widely.  Estimates do not need to come
  from \emph{studies}, these could also originate from different types
  of evidence, for example, from two meta-analyses, or from a
  meta-analysis and a single study.

  If the NNHM is fitted to the results of meta-analyses, then this
  adds another hierarchical level to the model.  In the spirit of a
  \emph{bias allowance model} framework \cite{WeltonEtAl}, in addition
  to between-study variability, the variability \emph{between study
    types} is considered as a separate variance component.
  Especially in the context of normal
  models \cite{BurkeEnsorRiley2017,MorrisEtAl2018} and when interest is
  in main effects \cite{Kontopantelis2018}, application of a one-stage
  model simultaneously including all hierarchy levels may in many
  cases not lead to substantially different results from a simpler
  two-stage approach in which data at the study-level are combined
  first, and summaries are subsequently combined in a second stage
  \cite{StewartTierney2002}. This way, inference is substantially
  simplified, and standard meta-analysis
  software can be used.
  
  Consider the example of a meta-analysis investigating the effect of
  immunosuppression in paediatric patients, where the outcome of
  interest is the occurrence of acute rejection (AR) events that the
  therapy is supposed to prevent \cite{CrinsEtAl2014}. Only two
  randomized trials are available, but in addition four observational
  studies reported on the effect. One may not expect to see identical
  effects in both types of studies, but the discrepancy between them
  will be limited. A meta-analysis of the two randomized trials may
  then profit from considering the outcomes of the four observational
  trials in addition, leading to a particular kind of
    extrapolation approach \cite{RoeverWandelFriede2018}.

\begin{figure}[t]
    \begin{center}
      \includegraphics[width=0.98\linewidth]{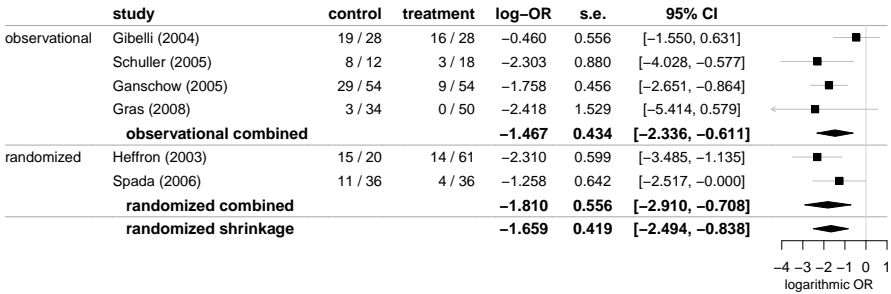} 
      \caption{\label{fig:crins-example} Illustration of a more
        general shrinkage application. The two sources of evidence
        themselves here are meta-analyses of observational and randomized
        studies. The combined randomized estimate may then borrow
        information from the observational studies' evidence. The two
        combined estimates are again meta-analysed to yield a
        shrinkage interval for the randomized effect.}
    \end{center}
  \end{figure}

  Figure~\ref{fig:crins-example} shows the example data. In both sets
  of studies we see similar effects, the negative combined estimates
  of the log-odds-ratio indicate a successful prevention of AR~events,
  and the two associated credible intervals are mostly overlapping.
  After combining the two sets of studies separately, we may now
  perform a meta-analysis of the resulting two combined estimates (in
  all cases using uniform priors for effects and HN($0.5$) priors for
  heterogeneities). The \emph{shrinkage} estimate for the mean effect in the randomized
  studies then provides an estimate for the randomized effect that is
  also informed by the observational evidence, while allowing for
  heterogeneity (at a second level) between both types of estimates.
  Note that in this context the shrinkage estimate then does not refer to a single study, but to one of the meta-analysis estimates that are combined here.
  The shrinkage estimate is shown at the very bottom of
  Figure~\ref{fig:crins-example}. Compared to the original estimate
  based only on the two randomized trials, the shrinkage estimate is,
  in concordance with the observational evidence, slightly more
  moderate (at a lower absolute log-OR).  Consideration of the
  additional evidence also gains precision: the shrinkage interval is
  $25\%$ shorter than the original interval.

  For the shrinkage estimate, we get a posterior probability of
  $\prob(\theta_\mathrm{rand.} \!>\!  0\,|\,y)=0.00007$ of a
  non-beneficial effect. With $p=0.0002$, the associated posterior
  predictive $p$-value again is of a similar magnitude. Compared to
  the original meta-analysis of $2$~randomized studies only, we can
  again see the gain in precision; here the evidence for a beneficial
  effect was not yet quite as pronounced
  ($\prob(\mu\!>\!0\,|\,y)=0.0023$ and $p=0.0079$).  
  The \textsf{R}~code to reproduce the calculations for this
    example is provided in the appendix.

\section{Discussion} \label{sec:discussion}
  Use of the NNHM to consider external information via shrinkage
  estimation provides a transparent procedure based on well-defined
  parameters and a common model framework. The NNHM may readily be
  generalized e.g. to more studies, more levels of hierarchy, or the
  inclusion of regression parameters. The amount of information
  considered may be explicated by noting that a joint analysis is
  equivalent to the use of a meta-analytic-predictive (MAP) prior
  \cite{SchmidliEtAl2014}.  At the same time, heavy tails of the
  MAP prior ensure a certain degree of robustness of the shrinkage
  estimate in case of prior-data conflicts \cite{OHaganPericchi2012}.
  The simulations demonstrate that the gain in precision may be
%
  \new{greater than expected, and substantial}
  especially in cases where the external data are
  associated with equal or less uncertainty than the data that are of
  primary interest.  The possible precision gain may allow the conduct
  and evaluation of trials in circumstances where otherwise evidence
  would be too sparse, or it may generally enable to allocate
  resources more efficiently.

  In the spirit of the \emph{reference model} parametrization outlined
  above, the institution of an ``overall mean''~$\mu$ is not
  necessary.  In many cases, when the data to be synthesized are of
  differing natures, the idea of a ``central'' mean parameter might be
  hard to motivate; what is relevant here is that the two estimates
  are modeled as being connected via an uncertain normally distributed
  offset. Normality here especially implies symmetry, i.e., the
  displacement between the two does not have a preferred direction;
  over- or under-estimation of one another are equally likely, so that
  \emph{a~priori} no systematic bias is assumed.
  \new{Availability of this alternative motivation broadens the range
    of applicability of meta-analytic methods.}

  As usual, the user needs to be aware of the limits of the
  applicability of the model, which here in particular means that the
  normality assumptions should be plausible
  \cite{JacksonWhite2018}. These assumptions might be challenged
  e.g. when estimates are based on count data suffering from
  small-sample or rare-event problems, in which case more specific
  models may be more appropriate \citep{JacksonEtAl2018}.  We also
  make the implicit assumption that patient populations are
  sufficiently similar to allow for a meaningful comparison.
  Furthermore, analyses of non-randomized studies may need to be
  adjusted for confounding
  \cite{SchmoorCaputoSchumacher2008,SignorovitchEtAl2010}, as was also
  done in the CJD example.

  Although frequentist analyses still dominate clinical trials,
  examples of Bayesian analyses are emerging. A recent application is
  the trial by Laptook \textit{et~al.} \cite{LaptookEtAl2017} in
  newborns with hypoxic-ischemic encephalopathy, a form of brain
  damage resulting from an insufficient supply of oxygen to the
  brain. The authors used the Bayesian framework to interpret their
  results in the light of different choices of priors that they termed
  ``neutral'', ``skeptical'' and
  ``optimistic'' \cite{QuintanaEtAl2017}. In this regard it differs
  from our proposal as we advocate the use of external data to inform
  the prior. 
  \new{The connection to common meta-analysis methods then helps motivating the choice of model details.}
  Sensitivity analyses could be performed in our setting by varying
  the prior on the between-trial heterogeneity $\tau$, e.g. by varying
  the scale parameter of the half-normal prior.

  Although not assessed in the simulations here, the performance of frequentist shrinkage BLUP estimators is likely to be unsatisfactory when dealing with only two studies. The reason lies in the underestimation of the between-study heterogeneity with a high likelihood of the variance estimate resulting in zero, and the challenge to incorporate the uncertaintly in the estimation of the heterogeneity in the inference \cite{BenderEtAl2018,FriedeRoeverWandelNeuenschwander2017a,FriedeRoeverWandelNeuenschwander2017b}. A Bayesian alternative was described here  and shown in simulations to have satisfactory properties under practically relevant scenarios. Therefore, the approach described here adds to the tool box of practicing statisticians.
%
  The proposed \new{Bayesian} approach can easily be implemented using the
  \textsf{R}~package \texttt{bayesmeta} \cite{bayesmeta,Roever2017} and
  relevant code is provided as appendices for the two-study and the
  two-meta-analyses cases.  The code to reproduce the simulations is also provided in the online supplement.
  The availability of posterior predictive
  $p$-values may aid in the interpretation of the findings.
  Furthermore, the efficient implementation facilitates sensitivity
  analyses and the assessment of operation characteristics of the
  procedures through simulations in so-called clinical scenario
  evaluations \cite{BendaEtAl2010}.

\begin{acks}
  This research has received funding from the EU's 7th Framework
  Programme for research, technological development and demonstration
  under grant agreement number FP HEALTH 2013-602144 with project
  title (acronym) ``Innovative methodology for small populations
  research'' (InSPiRe).
\end{acks}

\newpage
\section{Appendix}
\subsection{\textsf{R} code for CJD example}
{ \footnotesize 
\begin{verbatim}
# specify the data:
cjd <- cbind.data.frame("study"    = c("observational", "randomized"),
                        "logHR"    = c(-0.49948, -0.17344),
                        "logHR.se" = c(0.2493, 0.6312),
                        stringsAsFactors=FALSE)

# analyze:
require("bayesmeta")
bm <- bayesmeta(y         = cjd$logHR,
                sigma     = cjd$logHR.se,
                labels    = cjd$study,
                tau.prior = function(t){dhalfnormal(t, scale=0.5)})

# show results:
bm
forestplot(bm)

# show shrinkage estimates:
bm$theta

# interval length ratio (66%):
(q <- diff(bm$theta[7:8,"randomized"])
      / (2*qnorm(0.975)*bm$theta[2,"randomized"]))

# effective sample size gain (129%):
(1/q)^2 - 1

# heterogeneity prior median and 95% quantile:
qhalfnormal(c(0.50, 0.95), scale=0.5)
# heterogeneity posterior:
bm$summary[,"tau"]

# compute posterior predictive p-value
# (one-sided, for the randomized (shrinkage) effect,
#  and using the posterior probability of a beneficial effect
#  as the "test statistic"):
p1 <- pppvalue(bm, parameter="randomized", value=0,
               alternative="less", statistic="cdf",
               n=1000, seed=123)
p1

# for comparison, the posterior probability
# of a non-beneficial randomized effect: 
1 - bm$pposterior(theta=0, individual="randomized")
\end{verbatim} }

\newpage
\subsection{\textsf{R} code for paediatric transplantation example}
{ \footnotesize 
\begin{verbatim}
# load packages and data:
require("bayesmeta")
data("CrinsEtAl2014")

# compute effect sizes (log-OR for acute rejection (AR) events)
# using the metafor library's "escalc()" function;
# 4 observational studies:
effsize.obs <- escalc(ai = exp.AR.events,  n1i = exp.total,
                      ci = cont.AR.events, n2i = cont.total,
                      slab = publication, measure = "OR",
                      subset = (CrinsEtAl2014[,"randomized"]=="no"),
                      data = CrinsEtAl2014)
# 2 randomized studies:
effsize.rand <- escalc(ai = exp.AR.events,  n1i = exp.total,
                       ci = cont.AR.events, n2i = cont.total,
                       slab = publication, measure = "OR",
                       subset = (CrinsEtAl2014[,"randomized"]=="yes"),
                       data = CrinsEtAl2014)

# perform meta-analysis of 4 observational studies:
bm.obs <- bayesmeta(effsize.obs,
                    tau.prior = function(x){dhalfnormal(x,scale=0.5)})
# perform meta-analysis of 2 randomized studies:
bm.rand <- bayesmeta(effsize.rand,
                     tau.prior = function(x){dhalfnormal(x,scale=0.5)})

# perform 2nd-stage meta-analysis of previous MA results:
bm.combi <- bayesmeta(y      = c(bm.obs$summary["mean","mu"],
                                 bm.rand$summary["mean","mu"]),
                      sigma  = c(bm.obs$summary["sd","mu"],
                                 bm.rand$summary["sd","mu"]),
                      labels = c("observational", "randomized"),
                      tau.prior = function(x){dhalfnormal(x,scale=0.5)})

# compare "plain" randomized posterior and shrinkage estimate:
rbind("randomized-only" = bm.rand$summary[,"mu"],
      "shrinkage"       = bm.combi$theta[-(1:2),"randomized"])

# interval length ratio (75%):
(q <- diff(bm.combi$theta[7:8,"randomized"])
      / diff(bm.rand$summary[5:6,"mu"]))

# compute posterior predictive p-value:
p1 <- pppvalue(bm.combi, parameter="randomized", value=0,
               alternative="less", statistic="cdf",
               n=1000, seed=123)
p1
# posterior probability of a non-beneficial randomized effect: 
1 - bm.combi$pposterior(theta=0, individual="randomized")

# compute posterior predictive p-value
# for initial MA of 2 randomized studies only:
p2 <- pppvalue(bm.rand, parameter="mu", value=0,
               alternative="less", statistic="cdf",
               n=1000, seed=123)
p2
# original posterior probability of non-beneficial effect: 
1 - bm.rand$pposterior(mu=0)
\end{verbatim} }

\subsection{Model equivalence}
  Shrinkage estimation in the NNHM and in the \emph{reference model}
  introduced above yield identical results, as long as an improper
  uniform prior for the effect ($\mu$ or $\alpha$) is used. The
  heterogeneity prior densities are given by~$\tauprior(\tau)$ for
  then NNHM, and by
  $\frac{1}{\sqrt{2}}\tauprior\bigl(\frac{\beta}{\sqrt{2}}\bigr)$ for
  the reference model. Equivalence of the two models for the shrinkage
  estimates can be seen by comparing the resulting MAP priors
  $p(\theta_2|y_1)$ and $p(\vartheta_2|y_1)$.  To do so, we first
  introduce a reparametrisation (re-scaling) of the heterogeneity
  parameter as $\gamma=\textstyle\frac{\beta}{\sqrt{2}}$, where the
  new heterogeneity parameter's prior distribution then simply is
  given by~$\tauprior(\gamma)$.  The corresponding MAP prior densities
  then are given by
  \begin{eqnarray}
    p(\theta_2|y_1) 
    &=&
    \int \int p(\theta_2|\mu,\tau) \, p(\mu,\tau|y_1) \, \diff \mu \, \diff \tau \\
    & \propto &
    \int \biggl[\int p(\theta_2|\mu,\tau) \, p(y_1|\mu,\tau) \, \diff \mu \biggr] \, \tauprior(\tau)  \, \diff \tau
  \end{eqnarray}
  and 
  \begin{eqnarray}
    p(\vartheta_2|y_1)
    &=&
    \int \int p(\vartheta_2|\alpha,\gamma) \, p(\alpha,\gamma|y_1) \, \diff \alpha \, \diff \gamma \\
    & \propto &
    \int \biggl[\int p(\vartheta_2|\alpha,\gamma) \, p(y_1|\alpha,\gamma) \, \diff \alpha \biggr]\, \tauprior(\gamma)  \, \diff \gamma
  \end{eqnarray}
  In order to show that the integrals are identical, it now suffices
  to show that the terms in square brackets (the ``conditional MAP
  priors'' $p(\theta_2|y_1,\tau)$ and $p(\vartheta_2|y_1,\gamma)$,
  respectively) are identical.  For the NNHM we have
  \begin{eqnarray}
    & & \int p(\theta_2|\mu,\tau) \, p(y_1|\mu,\tau) \, \diff \mu \nonumber \\
    &=&
    \int {\textstyle\frac{1}{\sqrt{2\pi \tau^2}}} \,
    \exp\Bigl(\textstyle -\frac{1}{2} \frac{(\theta_2-\mu)^2}{\tau^2}\Bigr)
    {\textstyle\frac{1}{\sqrt{2\pi (\tau^2+\sigma_1^2)}}} \,
    \exp\Bigl(\textstyle -\frac{1}{2} \frac{(y_1-\mu)^2}{\tau^2+\sigma_1^2}\Bigr) 
    \; \diff \mu\\
    &=&
    {\textstyle\frac{1}{\sqrt{2\pi (2\tau^2+\sigma_1^2)}}} \,
    \exp\Bigl(\textstyle -\frac{1}{2} \frac{(\theta_2-y_1)^2}{2\tau^2+\sigma_1^2}\Bigr)
  \end{eqnarray}
  where the integral results as a convolution of two normal
  densities. Analogously, for the second variation we get
  \begin{eqnarray}
    & & \int p(\vartheta_2|\alpha,\gamma) \, p(y_1|\alpha,\gamma) \, \diff \alpha \nonumber \\
    &=&
    \int {\textstyle\frac{1}{\sqrt{2\pi\, 2 \gamma^2}}} \,
    \exp\Bigl(\textstyle -\frac{1}{2} \frac{(\alpha-\vartheta_2)^2}{2\gamma^2}\Bigr)
    \;
    {\textstyle\frac{1}{\sqrt{2\pi \sigma_1^2}}} 
    \exp\Bigl(\textstyle -\frac{1}{2} \frac{(\alpha-y_1)^2}{\sigma_1^2}\Bigr) 
    \;\diff\alpha \\
    &=&
    {\textstyle\frac{1}{\sqrt{2\pi (2\gamma^2+\sigma_1^2)}}} \,
    \exp\Bigl(\textstyle -\frac{1}{2} \frac{(\vartheta_2-y_1)^2}{2\gamma^2+\sigma_1^2}\Bigr)
  \end{eqnarray}
  With that, the two resulting MAP priors are identical, and the two
  models will yield the same results as far as the shrinkage
  estimates are concerned.

{ 
  \bibliographystyle{SageV}

}

\end{document}

%% file: shrinkCover.tex
\begin{table*}[ht]
\centering
\caption{\label{tab:shrinkCover}Coverage (\%) of shrinkage intervals
  for estimation of the first study's mean
  parameter~($\theta_1$). Sample sizes~($n_1$ and~$n_2$) as well as
  settings for the heterogeneity prior~($p(\tau)$) and actual
  heterogeneity values~($\tau$) are varied. The columns labelled by an
  asterisk~($\ast$) correspond to drawing the heterogeneity from its
  corresponding prior distribution.}\footnotesize
\begin{tabular}{lcccccccccccccc}
  \toprule
  \multicolumn{1}{r}{$\tau$~prior:} & \multicolumn{7}{c}{$\mathrm{HN}(0.5)$} & \multicolumn{7}{c}{$\mathrm{HN}(1.0)$} \\
  \cmidrule(lr){2-8} \cmidrule(lr){9-15}
  $n_1$/$n_2$ $\qquad\tau$: & 0.0 & 0.1 & 0.2 & 0.5 & 1.0 & 2.0 & $\ast$ & 0.0 & 0.1 & 0.2 & 0.5 & 1.0 & 2.0 & $\ast$ \\ 
  \midrule
  \vspace{1ex} 25/400 & 99.7 & 99.6 & 98.9 & 93.4 & 84.0 & 79.0 & 94.7 & 99.3 & 99.3 & 99.0 & 96.7 & 92.5 & 90.5 & 95.1 \\ 
   25/100 & 98.7 & 98.7 & 98.1 & 93.9 & 86.1 & 80.0 & 95.1 & 98.4 & 98.6 & 98.5 & 96.5 & 93.2 & 90.8 & 94.4 \\ 
  \vspace{1ex} 100/400 & 98.7 & 98.2 & 97.1 & 93.2 & 90.9 & 90.4 & 94.9 & 98.1 & 97.7 & 97.2 & 94.8 & 93.7 & 93.5 & 95.3 \\ 
   25/25 & 96.6 & 96.7 & 96.1 & 94.5 & 90.5 & 84.6 & 95.0 & 97.0 & 97.2 & 96.6 & 95.7 & 94.0 & 92.1 & 94.9 \\ 
   100/100 & 96.7 & 96.5 & 96.3 & 94.0 & 91.1 & 90.7 & 95.7 & 96.7 & 96.4 & 96.6 & 95.3 & 93.7 & 93.6 & 94.9 \\ 
  \vspace{1ex} 400/400 & 96.7 & 96.6 & 95.0 & 94.0 & 94.0 & 93.9 & 95.0 & 96.4 & 96.4 & 95.0 & 94.9 & 94.9 & 94.8 & 95.0 \\ 
   100/25 & 96.0 & 95.6 & 95.3 & 94.8 & 93.8 & 92.3 & 94.7 & 96.0 & 95.8 & 95.6 & 95.2 & 94.7 & 94.3 & 94.8 \\ 
  \vspace{1ex} 400/100 & 95.5 & 95.6 & 95.4 & 94.7 & 93.7 & 93.8 & 95.1 & 95.6 & 95.5 & 95.5 & 94.9 & 94.3 & 94.5 & 95.1 \\ 
   400/25 & 95.1 & 95.1 & 95.2 & 94.7 & 94.9 & 94.5 & 95.3 & 95.0 & 95.2 & 95.2 & 94.8 & 95.0 & 95.0 & 95.2 \\ 
   \bottomrule
\end{tabular}
\end{table*}

%% file: shrinkWidth.tex
\begin{table*}[ht]
\centering
\caption{\label{tab:shrinkWidth}Mean width (\%) of shrinkage intervals (for~$\theta_1$) relative to original ``plain'' CI based only on $y_1$ and $\sigma_1$.} \footnotesize
\begin{tabular}{lcccccccccccccc}
  \toprule
  \multicolumn{1}{r}{$\tau$~prior:} & \multicolumn{7}{c}{$\mathrm{HN}(0.5)$} & \multicolumn{7}{c}{$\mathrm{HN}(1.0)$} \\
  \cmidrule(lr){2-8} \cmidrule(lr){9-15}
  $n_1$/$n_2$ $\qquad\tau$: & 0.0 & 0.1 & 0.2 & 0.5 & 1.0 & 2.0 & $\ast$ & 0.0 & 0.1 & 0.2 & 0.5 & 1.0 & 2.0 & $\ast$ \\ 
  \midrule
  \vspace{1ex} 25/400 & 62.4 & 62.7 & 63.0 & 65.6 & 72.1 & 82.9 & 65.1 & 75.6 & 75.9 & 76.2 & 78.6 & 83.8 & 90.8 & 81.5 \\ 
   25/100 & 67.5 & 67.5 & 67.9 & 69.9 & 75.2 & 84.3 & 69.5 & 78.4 & 78.4 & 78.8 & 80.9 & 85.2 & 91.4 & 83.2 \\ 
  \vspace{1ex} 100/400 & 78.5 & 78.7 & 79.9 & 85.2 & 91.3 & 95.8 & 83.4 & 85.7 & 85.8 & 86.8 & 90.9 & 95.0 & 97.7 & 92.1 \\ 
   25/25 & 78.9 & 79.0 & 79.0 & 79.7 & 81.8 & 86.9 & 79.7 & 85.2 & 85.2 & 85.3 & 86.2 & 88.4 & 92.4 & 87.6 \\ 
   100/100 & 85.1 & 85.3 & 85.7 & 88.4 & 92.5 & 96.2 & 87.5 & 89.9 & 90.1 & 90.4 & 92.7 & 95.6 & 97.9 & 93.9 \\ 
  \vspace{1ex} 400/400 & 89.9 & 90.5 & 91.9 & 95.5 & 97.8 & 99.0 & 93.7 & 93.0 & 93.4 & 94.5 & 97.2 & 98.7 & 99.5 & 97.3 \\ 
   100/25 & 92.9 & 92.9 & 93.0 & 93.4 & 94.6 & 96.6 & 93.3 & 95.0 & 95.0 & 95.1 & 95.6 & 96.7 & 98.1 & 96.1 \\ 
  \vspace{1ex} 400/100 & 95.0 & 95.1 & 95.4 & 96.6 & 98.1 & 99.1 & 96.2 & 96.5 & 96.6 & 96.9 & 97.9 & 98.9 & 99.5 & 98.2 \\ 
   400/25 & 98.0 & 98.0 & 98.1 & 98.2 & 98.6 & 99.2 & 98.2 & 98.6 & 98.6 & 98.6 & 98.8 & 99.1 & 99.5 & 99.0 \\ 
   \bottomrule
\end{tabular}
\end{table*}

%% file: shrinkGain.tex
\begin{table*}[ht]
\centering
\caption{\label{tab:shrinkGain}Gain (\%) in effective sample size when using the shrinkage estimate, relative to the original CI.} \footnotesize
\begin{tabular}{lcccccccccccccc}
  \toprule
  \multicolumn{1}{r}{$\tau$~prior:} & \multicolumn{7}{c}{$\mathrm{HN}(0.5)$} & \multicolumn{7}{c}{$\mathrm{HN}(1.0)$} \\
  \cmidrule(lr){2-8} \cmidrule(lr){9-15}
  $n_1$/$n_2$ $\qquad\tau$: & 0.0 & 0.1 & 0.2 & 0.5 & 1.0 & 2.0 & $\ast$ & 0.0 & 0.1 & 0.2 & 0.5 & 1.0 & 2.0 & $\ast$ \\ 
  \midrule
  \vspace{1ex} 25/400 & 162.7 & 160.7 & 158.9 & 144.3 & 113.4 & 68.8 & 147.9 & 77.7 & 76.5 & 75.4 & 67.1 & 50.5 & 28.9 & 58.3 \\ 
   25/100 & 123.3 & 123.3 & 121.3 & 111.3 & 89.7 & 56.2 & 113.8 & 64.9 & 64.9 & 63.6 & 56.9 & 43.6 & 25.5 & 50.0 \\ 
  \vspace{1ex} 100/400 & 64.6 & 64.1 & 60.1 & 43.7 & 25.9 & 12.8 & 49.4 & 37.5 & 37.2 & 34.4 & 23.8 & 13.4 & 6.3 & 20.7 \\ 
   25/25 & 61.2 & 60.9 & 60.8 & 58.4 & 51.8 & 36.9 & 58.7 & 38.7 & 38.5 & 38.2 & 35.8 & 30.0 & 19.6 & 32.2 \\ 
   100/100 & 38.8 & 38.2 & 37.1 & 29.8 & 19.4 & 10.0 & 32.3 & 24.4 & 23.8 & 23.0 & 17.5 & 10.7 & 5.3 & 14.8 \\ 
  \vspace{1ex} 400/400 & 24.3 & 22.8 & 19.5 & 10.9 & 5.3 & 2.5 & 15.1 & 16.1 & 15.0 & 12.5 & 6.5 & 3.0 & 1.3 & 6.3 \\ 
   100/25 & 15.9 & 16.0 & 15.8 & 14.8 & 11.9 & 7.6 & 14.9 & 10.9 & 10.9 & 10.7 & 9.6 & 7.2 & 4.2 & 8.4 \\ 
  \vspace{1ex} 400/100 & 10.9 & 10.7 & 10.0 & 7.3 & 4.2 & 2.0 & 8.3 & 7.4 & 7.2 & 6.6 & 4.5 & 2.5 & 1.1 & 3.9 \\ 
   400/25 & 4.1 & 4.1 & 4.0 & 3.7 & 2.9 & 1.7 & 3.7 & 2.9 & 2.8 & 2.8 & 2.4 & 1.8 & 1.0 & 2.1 \\ 
   \bottomrule
\end{tabular}
\end{table*}

%% file: shortFraction.tex
\begin{table*}[ht]
\centering
\caption{\label{tab:shortFraction}Fraction (\%) of shrinkage intervals turning out shorter than the original CI.
         Note the differing ordering of rows compared to Tables~\ref{tab:shrinkCover}--\ref{tab:shrinkGain}.} \footnotesize
\begin{tabular}{lcccccccccccccc}
  \toprule
  \multicolumn{1}{r}{$\tau$~prior:} & \multicolumn{7}{c}{$\mathrm{HN}(0.5)$} & \multicolumn{7}{c}{$\mathrm{HN}(1.0)$} \\
  \cmidrule(lr){2-8} \cmidrule(lr){9-15}
  $n_1$/$n_2$ $\qquad\tau$: & 0.0 & 0.1 & 0.2 & 0.5 & 1.0 & 2.0 & $\ast$ & 0.0 & 0.1 & 0.2 & 0.5 & 1.0 & 2.0 & $\ast$ \\ 
  \midrule
   25/25 & 100.0 & 99.9 & 100.0 & 99.7 & 97.4 & 81.6 & 99.5 & 99.4 & 99.2 & 99.1 & 97.8 & 91.1 & 68.6 & 91.4 \\ 
   25/100 & 99.9 & 99.9 & 99.9 & 99.1 & 92.3 & 68.7 & 98.6 & 99.2 & 99.2 & 98.8 & 96.1 & 83.9 & 57.6 & 86.9 \\ 
  \vspace{1ex} 25/400 & 99.9 & 99.9 & 99.9 & 98.9 & 90.7 & 64.5 & 98.1 & 99.3 & 99.3 & 99.1 & 95.8 & 82.1 & 54.4 & 85.8 \\ 
   100/25 & 99.8 & 99.8 & 99.7 & 98.6 & 90.0 & 65.7 & 98.1 & 98.3 & 98.2 & 97.9 & 94.5 & 80.4 & 54.1 & 85.2 \\ 
   100/100 & 99.4 & 99.0 & 98.5 & 91.0 & 68.8 & 39.6 & 91.5 & 97.6 & 96.8 & 95.5 & 83.8 & 59.8 & 33.4 & 71.3 \\ 
  \vspace{1ex} 100/400 & 99.1 & 98.7 & 97.4 & 84.2 & 57.0 & 31.4 & 87.2 & 97.4 & 96.9 & 94.5 & 77.0 & 50.1 & 27.0 & 65.4 \\ 
   400/25 & 99.7 & 99.8 & 99.5 & 97.6 & 86.9 & 59.3 & 96.9 & 98.1 & 98.1 & 97.0 & 92.8 & 76.2 & 48.8 & 82.3 \\ 
   400/100 & 98.6 & 98.1 & 95.8 & 80.6 & 54.4 & 29.2 & 84.7 & 96.1 & 95.0 & 91.2 & 72.8 & 46.9 & 24.5 & 62.3 \\ 
   400/400 & 97.6 & 95.6 & 88.5 & 60.1 & 33.7 & 17.7 & 72.0 & 95.0 & 92.1 & 83.2 & 54.1 & 30.0 & 15.5 & 48.6 \\ 
   \bottomrule
\end{tabular}
\end{table*}